\documentclass[conference]{IEEEtran}
\usepackage{graphicx}
\usepackage{caption}
\usepackage{subcaption}
\usepackage{amsmath}
\usepackage{soul,color}
\usepackage{todonotes}
\usepackage{url}
\usepackage{mathtools}
\usepackage{calc}
\usepackage{xparse}
\usepackage{algorithm}
\usepackage{algpseudocode}
\usepackage{pifont}
\usepackage{makeidx}

\newcounter{AFNumberOfComments}
\stepcounter{AFNumberOfComments}
\newcounter{MDNumberOfComments}
\stepcounter{MDNumberOfComments}
\newcounter{MHNumberOfComments}
\stepcounter{MHNumberOfComments}

\begin{document}

\title{Service on Demand: Drone Base Stations\\ Cruising in the Cellular Network }

\author{\IEEEauthorblockN{Azade Fotouhi\IEEEauthorrefmark{1}\IEEEauthorrefmark{2},
Ming Ding\IEEEauthorrefmark{2} and
Mahbub Hassan\IEEEauthorrefmark{1}\IEEEauthorrefmark{2}}
\IEEEauthorblockA{\IEEEauthorrefmark{1}
University of New South Wales (UNSW), Sydney, Australia \{a.fotouhi, mahbub.hassan\}@unsw.edu.au}
\IEEEauthorblockA{\IEEEauthorrefmark{2}CSIRO Data61, Australia \{azade.fotouhi, ming.ding, mahbub.hassan\}@data61.csiro.au}
}

\maketitle

\begin{abstract}
In this paper, the deployment of drone base stations to provide higher performance in the cellular networks is analyzed. In particular, we investigate a new mobility model for drone base stations where they can move freely in the network, ignoring the cell boundaries. Free movement model for drones bring out new challenges such as user association and physical collision among drones. We consider two user association schemes and evaluate their performance through simulation. We show that by deploying a smart user association scheme in the free movement model, the obtained results are greatly better than those restricting each drone base station to fly over a certain small cell area, and serving local users. Additionally, the impact of drones' movement on the load balance, signal strength and interference are studied. Moreover, we show that our proposed algorithm can maintain a comfortable distance among the drones to avoid physical collision.

\end{abstract}

\IEEEpeerreviewmaketitle

\section{Introduction} \label{sec:intro}

Drones or small unmanned aerial vehicles (UAVs) are becoming a promising solution for a wide range of civilian applications such as disaster recovery, traffic monitoring, and surveillance. Due to their high degree of freedom, and the ability to move autonomously to any hard-to-reach- areas, they are emerging in cellular network applications as well to provide coverage and higher quality services for the users. Drones can be equipped with the base station (BS) hardware and act as a flying BS, creating an attractive alternative to conventional roof or pole mounted base stations.
The concept of drone base station (DBS) is still in its infancy, and many academic researchers are now actively working in the area.

While recent studies \cite{al2014optimal,yaliniz2016efficient} on DBS mainly focused on finding the optimum location for the drones to hover so that the coverage is maximized, we are utilizing the flexibility and agility of drones and study the DBSs that can move continuously over the serving area. DBSs can adapt their directions in order to provide higher service quality for the mobile users, moving randomly within the small cell boundaries.

In our previous works \cite{tmc_submitted2017,wowmom_main2017}, we designed drone mobility control algorithms according to drone's practical limitation \cite{Shanmugavel20101084}, in order to improve the performance of the cellular network. In the network area, divided to multiple small cell area, each DBS' mobility was limited to its small cell boundaries. All users in the small cell were remained to be associated to their local DBS all the time; although during movement of users, they might find another DBS with a higher received signal strength. We have shown that letting drones chasing users can significantly improve the system performance, especially the packet throughput for cell-edge users.

Now, an intriguing question is: how about further freeing the drones and allowing them to fly over the entire network, instead of over a single cell. Our motivation is to increase the DBS mobility range, thus providing more candidate DBSs for users to connect with.

However, the free movement model inherently requires the change of user association scheme. In more detail, due to the free movement of DBSs, a user may frequently find different DBSs available for communication. Therefore, users should be able to reselect their serving DBS in the network area. We consider two different user association schemes in this paper. We first show that a simple user association scheme only based on the received signal strength does not yield performance benefits in terms of system throughput from the free movement model because the traffic load is unbalanced in the network, e.g., a DBS may serve many users due to a good geometry condition while another DBS may be idle.

To restore the load balance as achieved by the restricted movement model, we propose a more advanced user association scheme that jointly considers the signal strength and the load among UAVs.


The rest of the paper is structured as follows.
The system model is presented in Section \ref{sec:systemmodel},
followed by performance metrics in Section \ref{sec:performancemetrics}. Our proposed user association schemes are presented in Section \ref{sec:userassoc}. We then review our proposed drone movement algorithm in Section \ref{sec:proposedalg}.
In Section \ref{sec:simulation}, the simulation results are presented.
Finally the conclusion and future work are discussed in Section~\ref{sec:conclusion}.

\section{System Model} \label{sec:systemmodel}
\subsection{Network Scenario}
Assume there is a large network area in size of $L(m) \times L(m)$, to be covered by drone base stations flying above the area.
The target area is divided to $C$ small cells, each of size $l(m) \times l(m)$. In each small cell area $U_s$ users are moving according to Random Way Point model (RWP). In this model, each user selects a random destination within the small cell border independent of other users, and moves there following a straight trajectory with a constant speed selected randomly from a given range. Upon reaching the destination, users may pause for a while before continuing to move to another destination \cite{rwp1,rwp2}.
The total number of users in the network is equal to $U_m = U_s\times C$.

\begin{figure}
	\centering
	\includegraphics[scale=0.35]{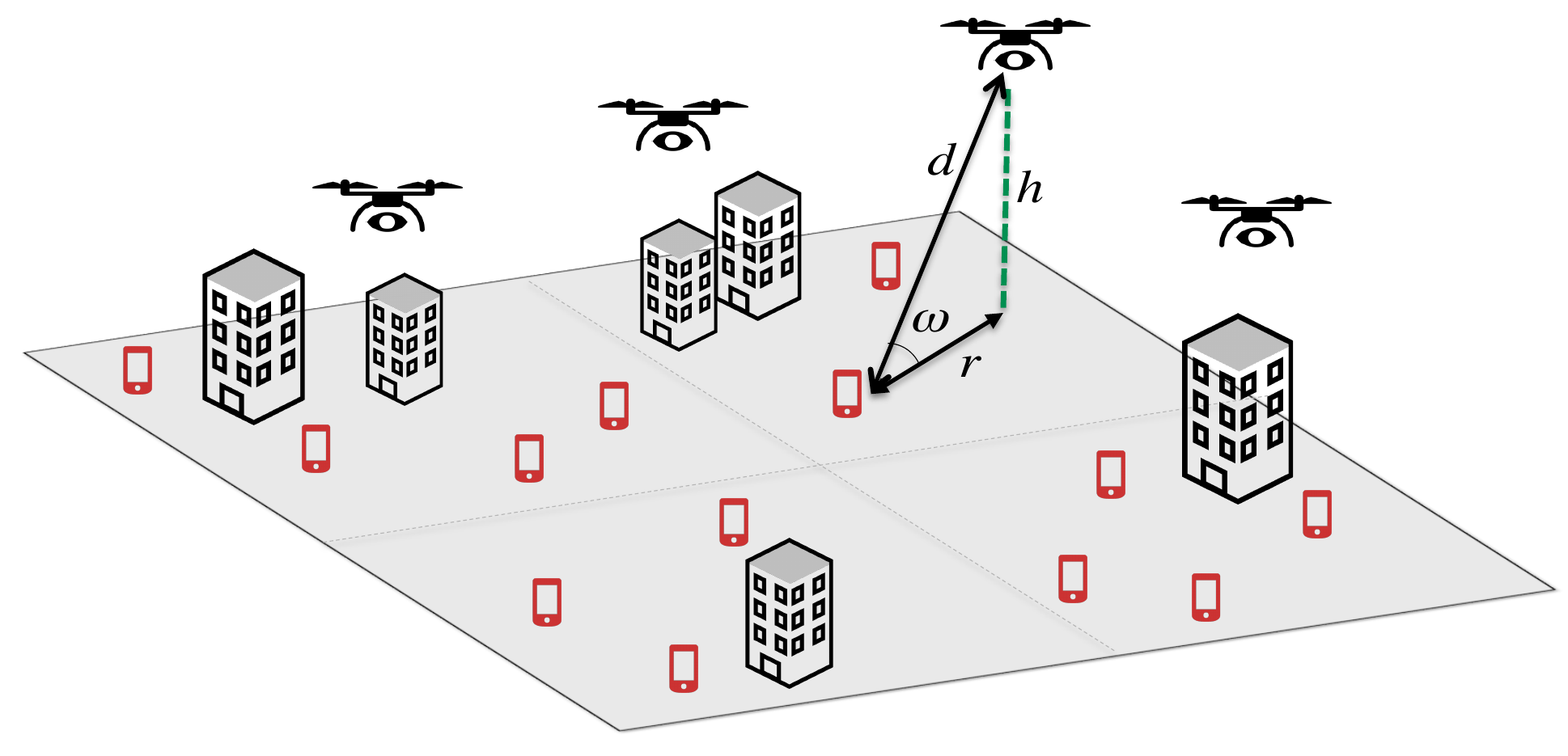}
	\caption{The network area with multiple mobile users and DBSs}
	\label{fig:arch}
\end{figure}

Moreover, there are $N$ drone base stations, constantly moving in the network with constant speed $v$ (m/s), at fixed altitude of $h$ (m). Figure \ref{fig:arch} shows the considered network architecture.

Note that, deploying drones at the same height with free movement would cause collision among drones. One alternative to avoid the collision issue is using height separation technique. However, by using height separation, drones could be deployed at a very wide range of height, causing performance degradation for the system. As a result, we install all DBSs at the same height, and then address the possibility of collision.

DBSs may be connected to a nearby cell tower with a wireless backhaul link.
We further assume that each DBS is transmitting data to users using a fixed transmission power of $p_{tx}$ (watt), total bandwidth $B$ (Hz) with central carrier frequency of $f$ (Hz). It is assumed that transmission from a DBS can create interference on mobile users in the serving area up to $\kappa$ meter. The interference beyond $\kappa$ meter is negligible.

The \textit{ground distance} or the two-dimensional (2D) distance between user $u~( u \in [1,2,\dots,U_m])$, and drone $n~(n \in [1,2,\dots,N])$ is defined by the distance between the user and the projection of the drone location onto the ground, denoted by $r_{u,n}$. The \textit{euclidean distance} or the three-dimensional (3D) distance between user $u$ and drone $n$ is presented by $d_{u,n} = \sqrt{r_{u,n}^2 +h^2}$, where $h$ is the height of drones.

\subsection{Channel Model} \label{sec:channel}

In this paper,
we consider a practical path loss model incorporating both LoS (Line of Sight), and NLoS (Non Line of Sight) transmissions.
More specifically,
the path loss function is formulated according to a probabilistic LoS model~\cite{al2014optimal,7194055},
in which the probability of having a LoS connection between a drone and its user depends on the elevation angle of the transmission link.
According to~\cite{al2014optimal},
the LoS probability function is expressed as
\begin{equation}
P^{LoS}(u,n) = \frac{1}{1+\alpha exp(-\beta[\omega -\alpha])},
\label{eq:plos}
\end{equation}
where $\alpha$ and $\beta$ are environment-dependent constants,
$\omega$ equals to $arctan(h/r_{u,n})$ in degree.
As a result of (\ref{eq:plos}),
the probability of having a NLoS connection can be written as
\begin{equation}
P^{NLoS}(u,n) = 1 - P^{LoS}(u,n).
\label{eq:pnlos}
\end{equation}

From (\ref{eq:plos}) and (\ref{eq:pnlos}), the path loss in dB can be modeled as
\begin{equation}
\eta_{path}(u,n) = A_{path} + 10\gamma_{path}\log_{10}(d_{u,n}),
\label{eq:pathloss}
\end{equation}
where the string variable \textit{``path"} takes the value of {``LoS"} and {``NLoS"} for the LoS and the NLoS cases, respectively.
In addition,
$A_{path}$ is the path loss at the reference distance (1 meter) and $\gamma_{path}$ is the path loss exponent,
both obtainable from field tests \cite{TR36.828}.

\subsection{Traffic Model}\label{sec:traffic_model}

The traffic model for each user follows the recommended traffic model by 3GPP \cite{3gpp36814}. In this model, there is a reading time interval between two subsequent user's data packet request. The reading time of each data packet is modeled as an exponential distribution with a mean of $\lambda$ (sec). Moreover, the transmission time for each data packet is defined as the time interval between the request time of a data packet and the end of its download, denoted by $\tau$ (sec).


All data packets are assumed to have a fixed size of $p$ (MByte). The user is called an \textit{active} user during the transmission time.

\subsection{Drone Mobility Control } \label{subsec:dronemobility}
All DBSs have the same height, therefore we consider their mobility in the 2D plane only. Each drone moves \textit{continuously} in the 2D space with a constant linear speed of $v$, and updates its moving direction every $t_{m}$ sec, hereafter called \textit{Direction Update Interval}. The proposed continuously moving model is thus applicable to all types of drones, with or without rotors.


When the drone wants to change its direction while keeping a constant speed, it moves along an arc. More importantly, the maximum possible turning angle $\theta_{max}$ for a drone during a specific time $t_m$  can be obtained by $\theta_{max} = \displaystyle \frac{a_{max} \times t_m}{v}$, where $ a_{max}$ and $v$ is the maximum acceleration and the speed of drone, respectively \cite{Shanmugavel20101084,Agility1998}.
At every $t_m$, the DBS chooses an angle, $\theta_n$, between $\pm$[0,$ \theta_{max} $]
and starts to complete the turn at the end of next $t_m$ sec.

\section{Performance Metrics} \label{sec:performancemetrics}

The main motivation for the proposed model is to improve the system capacity. In this section, we define the required metrics to evaluate the network performance.

The received signal power, $S^{path}(u,n)$ (watt), of an active user $u$ associated to drone $n$ can be obtained by

\begin{align}
\begin{split}
S^{path}(u,n)&=\frac{b_u}{B} \times p_{tx} \times 10^{\frac{-\eta_{path}(u,n)}{10}}
\end{split}
\label{eq:rcvpower}
\end{align}
where $b_u~(0 \leq b_u \leq B) $ is the allocated bandwidth to the user.

Moreover, the total noise power, $N_u$ (watt), for an active user $u$ including the thermal noise power and the user equipment noise figure,
can be represented by \cite{thermalnoise}
\begin{equation}
N_u = 10^{ \frac{-174+\delta_{ue}}{10}}\times{b_u}\times10^{-3},
\label{eq:noise}
\end{equation}
where $\delta_{ue}$ (dB) is the user equipment noise figure.

Accordingly, the \textit{Signal to Noise (SNR)} and \textit{Signal to Interference plus Noise Ratio (SINR)} of user $u$ associated to drone $n$ can be expressed as:
\begin{align}
\begin{split}
SNR^{path}(u,n)=\frac{S^{path}(u,n)}{N_u},
\end{split}
\label{eq:snr}
\end{align}
\begin{align}
\begin{split}
SINR^{path}(u,n)&=\frac{S^{path}(u,n)}{I_u+N_u},
\end{split}
\label{eq:sinr}
\end{align}
where $I_u = \big(\sum_{i \in {N}, i \not= n, r_{u,i} \leq \kappa } S^{path}(u,i)\big)$ represents the interference signal from neighbor DBSs received by user $u$.

Then, the \textit{spectral efficiency (SE)} (bps/Hz) of an active user $u$ associated with drone $n$ can be formulated according to the Shannon Capacity Theorem as~\cite{Book_Proakis}
\begin{align}
\begin{split}
\Phi^{path}(u,n) = \log_2 (1+SINR^{path}(u,n)).
\end{split}
\label{eq:individualspec}
\end{align}

Given the probabilistic channel model,
the average SE for user $u$ can be expressed as
\begin{align}
\begin{split}
\bar{\Phi}(u,n) =P^{LoS}\times\Phi^{LoS}(u,n)
+ P^{NLoS}\times\Phi^{NLoS}(u,n).
\end{split}
\label{eq:averagespec}
\end{align}

Moreover, the \textit{Throughput} (bps) of an communication link between an active user $u$ and drone $n$ can be formulates as
\begin{align}
\begin{split}
T(u,n) = b_u\times\bar{\Phi}(u,n).
\end{split}
\label{eq:throughput}
\end{align}

Additionally, \textit{Packet Throughput}, the ratio of successfully transmitted bits over the time consumed to transmit the said data bits, can be expressed as
\begin{align}
\begin{split}
P = p \times \frac{1}{\tau}.
\end{split}
\label{eq:packetthroughput}
\end{align}

Considering all downloaded packets by all users, the average packet throughout is considered as a
performance metric.

\section{User Association Schemes} \label{sec:userassoc}
At any specific time, a set of users are connected to a DBS, however, in the free movement models, users can reselect their serving DBSs frequently. The set of all active users associated to a DBS $n$ during at a specific time $t$ is denoted by ${\mathcal{Q}}_{n}(t)$.
Additionally,
the total bandwidth of $B$ is shared \textit{equally} among all associated active users of a DBS, and the DBS updates resource allocation every $t_r$ sec, called \textit{Resource Allocation Interval}.
In the following, two proposed schemes to control user association process are described.

\subsection{RSS-Based Scheme}
In this scheme, a user selects a DBS with the highest \textit{Received Signal Strength} (RSS), and can reselect its serving DBS every $t_r$. There is no limitation on the number of users that can be associated to a specific DBS. Note that each user can independently choose its serving DBS according to the observed RSS without any additional information from the other users.

\subsection{Throughput-Based Scheme}
By only taking into account the \textit{RSS}, a large number of users might select one DBS at the same time, thus creating unbalanced loads among DBSs and in turn reducing the system throughput due to the under-utilization of the frequency spectrum. To overcome this problem, we consider a more advanced association scheme, which needs global network knowledge.

In this model, a user selects a DBS that can maximize the estimated throughput for the next resource allocation interval. In particular, when a user $u'$ requests a new packet at time $t$, the system throughputs based on the hypotheses of its association with each candidate DBS in the network area is estimated for time $t'= t +t_r$. The DBS that gives the highest system throughput will be selected to serve $u'$. To solve this problem, we first define a binary association variable as follow
\[
x_i = \begin{dcases*}
1  & if DBS \textit{i} is selected\\
0 & otherwise
\end{dcases*}
\]
for $i \in N$. Then, the optimization problem to find the best DBS for user $u'$ can be expressed as
\begin{equation}
\max\limits_{ x_i \in \{0,1\}}\  \sum_{i=1}^{N} \Big(\sum_{u=1}^{\mathcal{Q}_{n}(t')} T(i,u) \Big)
\label{eq:th_assoc_objective}
\end{equation}
\begin{equation}
s.t. \quad  \mathcal{Q}_{n}(t') = \mathcal{Q}_{n}(t)+x_i(u') \quad\quad
\label{eq:constraint1}
\end{equation}
\begin{equation}
 \sum_{i=1}^{N} x_i  =1
 \label{eq:constraint2}
\end{equation}
The first constraint defines the set of associated users in each DBS considering serving/not serving the new request. To make sure that the user is connected to just one DBS, the second constraint must be satisfied.
\section{DBS Mobility Algorithms} \label{sec:proposedalg}

In our previous work \cite{tmc_submitted2017}, we proposed three different DBS mobility algorithms (DMAs). We showed that the one that employs Game Theory to make mobility decisions for DBSs performed the best. Therefore, in this paper we only consider the Game Theory based DMA.

The task of a DMA is to choose turning angles for DBSs at the start of every $t_m$ interval to improve the performance of the system.
The DBS will continue to follow the path specified by the turning angle selected at the \textit{start} of the interval for the next $t_m$ seconds.
This path cannot be changed in the middle of $t_m$ despite any further changes in mobile user population and traffic in the system.
When there is no associated user to a DBS, it chooses a random direction that keeps the drone in the intended border.

To reduce the complexity of the problem, we discretized all turning options into a finite set of $[-\theta_{max},\dots,-2g,-g,0,g,2g,\dots,\theta_{max}]$,
where $g=\displaystyle \frac{2\theta_{max}}{G-1}$, with $G$ representing the total number of turning options.
Each drone can choose its direction from $G$ candidate ones.

In the game theory based DMA, the direction selection is formulated as a non-cooperative game played by all serving DBSs in the system. The game is played at the start of each $t_m$ interval and the decisions leading to the Nash Equilibrium (NE) are adopted by the DBSs to update their directions. A pure NE is a convergence point where no player has an incentive to deviate from it by changing its action.
Hereafter, we refer to this algorithm as \textit{GT} DMA.

The game is described by ${\mathcal G} = ({\mathcal P},\{{\mathcal A}_p\},u_p)$, where ${\mathcal P} = \{ 1,2,\dots, P\}$ is the set of DBSs as players with at least one associated active user. ${\mathcal A}_p $ is the set of actions ($G$ turning angles) for each DBS, and $u_p$ is the utility function of each DBS.

Furthermore, $u_p:{\mathcal A}\rightarrow {\rm I\!R}  $ maps any member of the action space, $\theta \in {\mathcal A}$, to a numerical real number. The action space ${\mathcal A}$ is defined as the Cartesian product of the set of actions of all players (${\mathcal A} = {\mathcal A}_1 \times {\mathcal A}_2 \times \dots \times {\mathcal A}_P $). We denote the utility function of each player as $u_p(\theta_p, \theta_{-p})$, where $\theta_{-p}$ presents the action of all players except $p$. The utility function for each player is defined by the spectral efficiency of that player given the action of all players, as follows

\begin{align}
\begin{split}
u_p(\theta) = u_p(\theta_p, \theta_{-p}) =  \bar{ {\Phi}}(p),
\end{split}
\label{eq:utilityfunc_def}
\end{align}
where $ \bar{ {\Phi}}(p) $ is the average SE for the active users associated to DBS $p$.

In a non-cooperative game, each player independently tries to find an action that maximizes its own utility, however its decision is influenced by the action of other players:

\begin{align}
\begin{split}
\theta_{n} = \text{arg}\,\max\limits_{\forall  {\theta_{p}\in {\mathcal A}_{p}}}\ u_p(\theta_p, \theta_{-p})  \quad \forall p \in  \mathcal{P}.
\end{split}
\label{eq:maxutfun_game}
\end{align}

In this algorithm, at first, all drones select a random direction from their set of actions. Then each of them finds their best response considering other players' action.  Finally, after few trials they all converge to a NE point and move towards the selected directions during the next $t_m$ interval.

\section{Evaluation and Simulation Results} \label{sec:simulation}

In this section, the performance of the DBS network is evaluated through extensive simulation by MATLAB.
We refer to the model where drones are free to move in the network by a prefix of \textit{Free}. In this model, either a RSS-based or Throughput-based user association scheme can be employed.
On the other hand, the prefix of \textit{Restricted}, represents the models where users are always associated to their local DBSs which are restricted to move over the small cell areas. Finally, \textit{HOV} denotes the models where hovering DBSs are deployed over the center of the small cell areas.

\subsection{Simulation Setup}

The network area is divided into a 7$\times$7 grid of small cells (49 small cells), each of size $80m \times 80m$. Due to interference, \textit{outer} cells in the simulated network scenario will receive less interference than \textit{inner} cells. To obtain unbiased performance results, data is collected only from users in the inner cell. 

We used the same physical setting for the drones as our previous works \cite{wowmom_main2017,7848883,tmc_submitted2017}. The drones's speed vary from 2m/s to 8m/s, with the capability of changing direction every $t_m =1s$. Moreover, the current observed drone acceleration is set to 4 $m/s^2$ \cite{wowmom_ws2017}, while higher accelerations are expected for future drones.

The recommended height of 10m \cite{7842150} is selected for all DBSs in our simulation.
The number of users and their traffic model follow the parameters recommended by the 3GPP~\cite{3gpp36814}, and are shown in Table~\ref{tbl:params}.
Moreover,
to mitigate the randomness of the results,
all results have been averaged over 10 independent runs of 800-second simulations.

\begin{table}[]
	\centering
	\caption{Definition of parameters and their value}
	\label{tbl:params}
	\begin{tabular}{lll}
		\hline
		{\bf Symbol} & {\bf Definition} & {\bf Value}  	 \\
		\hline
		\hline
		$N$	& Number of Drones	& 49\\
		$C$ & Number of Small Cells & 49 \\
		$U_m$	&Total Number of Users in the Area&245\\
		$U_s$	&Number of Users in Each Small Cell&5\\
		$B$		& Total Bandwidth   &5 MHz\\
		$h$		& Drone Height         &10 m\\
		$v$		& Drone Speed		&[2, 4, 6, 8] m/s	\\
		$w$		& Edge Length of a Small Cell	&80m	\\
		$f$		&Working Frequency		&2 GHz		\\
		$p_{tx}$& Drone Transmission Power& 24 dBm \cite{TR36.828}\\
		$\lambda$		& Mean Reading Time	& 40 sec \\
		$\alpha, \beta$ & Environmental Parameters for Urban Area& 9.61, 0.16 \cite{yaliniz2016efficient}    \\
		$\gamma$	& Path Loss Exponent (LoS/NLoS)& 2.09/3.75 \cite{TR36.828}\\
		$\delta_{ue}$     & UE Noise Figure      & 9 dB      \\
		${t}_{m}$   &Direction Update Interval 	&1 sec \\
		$t_{r}$   &Resource Allocation Interval	&0.2 sec \\
		$\kappa$   &Interference Distance &  200 m\\
		$p$		&Data Size	& 4MByte		\\
		$G$   & Number of Candidate Directions&21  \\
		\hline
	\end{tabular}
\end{table}
\begin{figure}
	\centering
	\includegraphics[width=1\linewidth]{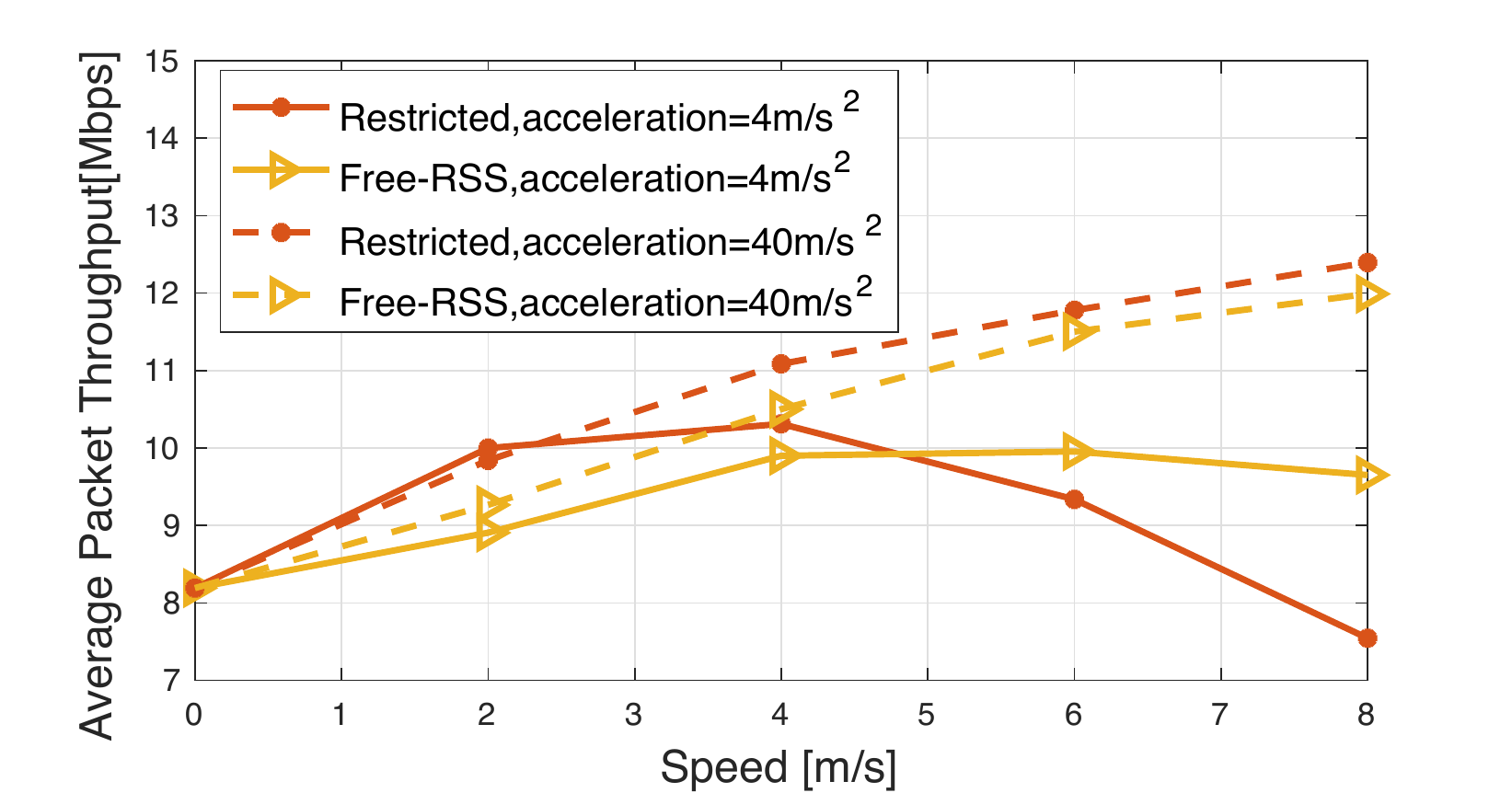}
	\caption{Average packet throughput for Restricted and Free-RSS movement model}
	\label{fig:avgpcktth}
\end{figure}
\begin{figure}
	\centering
	\includegraphics[scale=0.6]{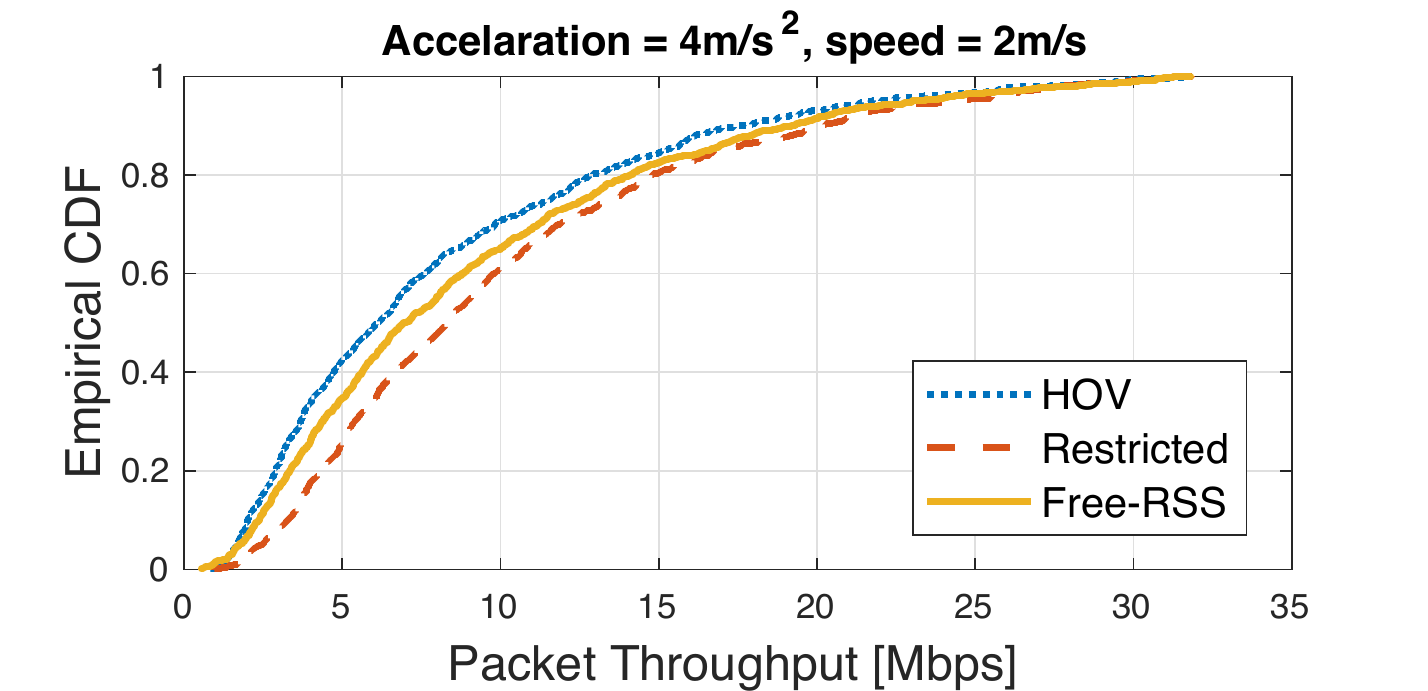}
	\caption{Empirical CDF of packet throughput for speed of 2m/s}
	\label{fig:packet_th}
\end{figure}
\begin{figure*}
	\centering
	\includegraphics[scale=0.54]{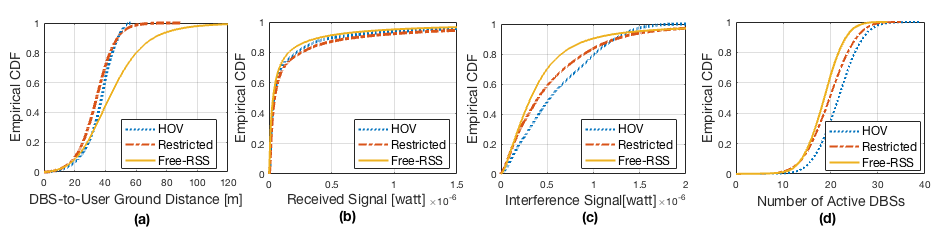}
	\caption{CDF of (a) DBS-to-user distance (b) Received Signal (c) Interference Signal and (d) Number of active DBSs}
	\label{fig:subplot_cdf}
\end{figure*}
\subsection{The Performance of RSS-based User Association} \label{subsec:performance_rslt}
In this section, we study the performance of the RSS-base user association with the free movement model.
Figure \ref{fig:avgpcktth} plots the average packet throughput of the system when DBSs are moving with various speeds, while the speed of ``0'' represents the \textit{HOV} scenario. From this figure, we can draw the following observations:
\begin{itemize}
	\item Generally speaking, the average packet throughput of the \textit{Free-RSS} movement model is lower than that of the \textit{Restricted} movement model, especially when DBSs are moving with a low speed. However, the \textit{Free-RSS} movement model still outperforms the HOV scenario. We further study the behavior of various metrics in the network to see how a lower packet throughput than that of \textit{Restricted} movement model is obtained in the \textit{Free-RSS} movement model.
	\item Similar to the observed results in our previous works, a higher acceleration generates better results than a lower acceleration.
\end{itemize}

In order to show why the restricted movement model yields a worse performance than the Free-RSS movement model, we focus on the scenario when DBSs are moving with the speed of 2m/s, with maximum acceleration of 4$m/s^2$.

First, the empirical CDF of the packet throughput is plotted in Figure \ref{fig:packet_th}.
It can be observed form this figure that \textit{Free-RSS} model outperforms the \textit{HOV} model in terms of packet throughout, however, the \textit{Restricted} model generates higher performance than the \textit{Free-RSS} model.

We further collected the ground distance between any active user and its associated DBS during the simulation time. The CDF of the ground distance is depicted in Figure \ref{fig:subplot_cdf}a. According to this figure, when drones follow the \textit{Free-RSS} movement model, the DBS-to-user distance becomes higher than both HOV and \textit{Restricted} movement model.
A larger DBS-to-user distance deteriorates the received signal strength, as illustrated in Figure \ref{fig:subplot_cdf}b. Users in \textit{Free-RSS} movement model receives a lower signal strength than that of \textit{HOV} and \textit{Restricted} scenarios.

We also investigate the interference signal at the users; as shown in Figure \ref{fig:subplot_cdf}c, the \textit{Free-RSS} movement model has the lowest interference signal in the network. It means that a less number of DBSs are transmitting data to users at a specific time, creating less interference. Figure \ref{fig:subplot_cdf}d confirms that the number of active DBSs during the simulation time in the free model movement is indeed less than other models.

Additionally, in the \textit{Free-RSS} movement model, the number of associated active users to a DBS might change over time. To see how the DBS loads vary during the simulation time, the number of active users associated with each DBS are collected for both the \textit{Free} and \textit{Restricted} movement models, and compared with the hovering drones.
\begin{figure}
	\centering
	\includegraphics[scale=0.5]{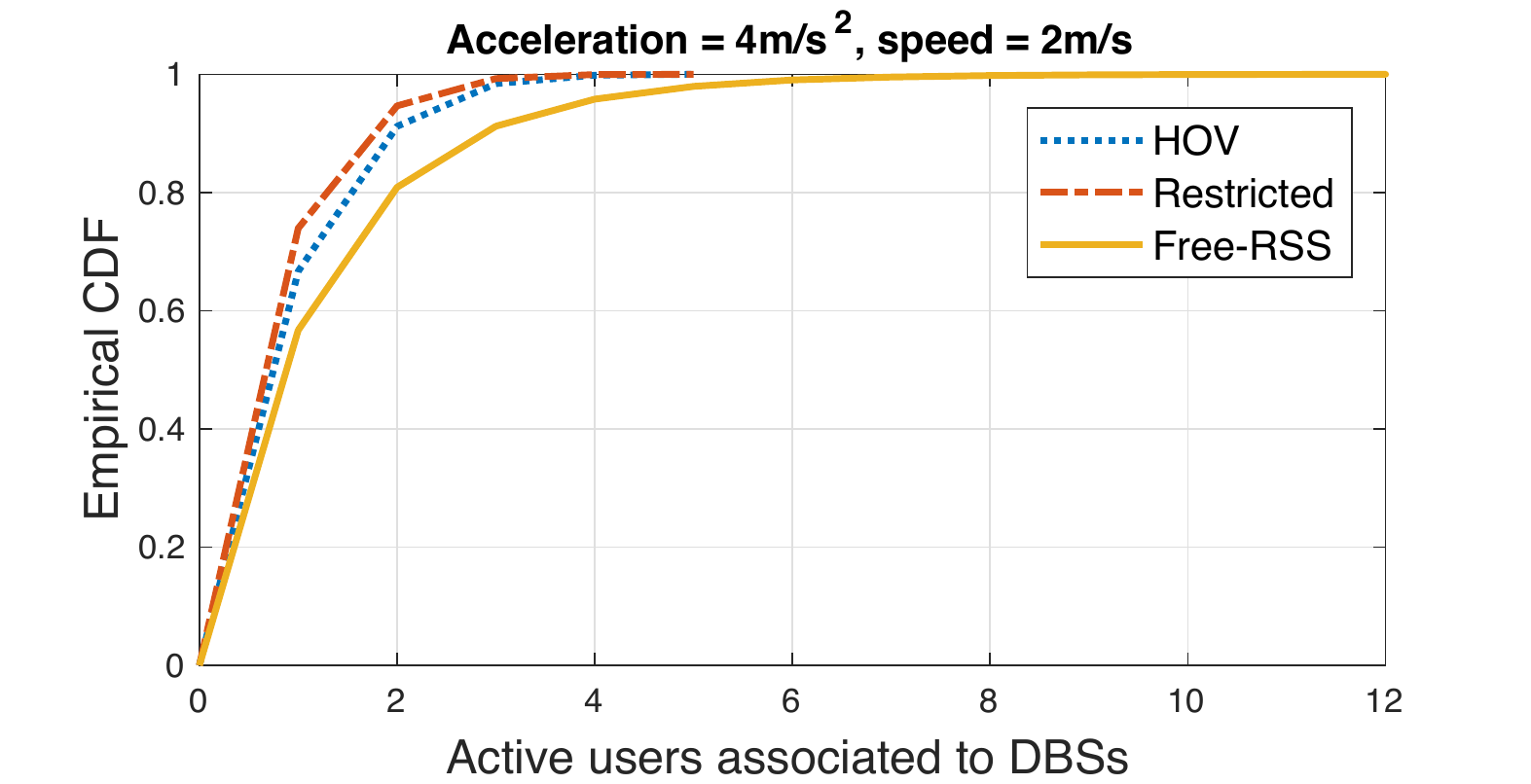}
	\caption{CDF of number of associated active users to DBSs }
	\label{fig:assoc_users}
\end{figure}
As can be seen from Figure \ref{fig:assoc_users}, there is a possibility of having a large number of users associated to one drone in \textit{Free-RSS} model, which causes unbalanced load among DBSs.
In contrast, in the restricted and HOV models, the maximum number of associated users to a DBS is fixed to the number of users in a small cell.
Note that unbalanced loads among DBSs reduces the
system throughput due to the under-utilization of the frequency
spectrum.
\subsection{The Performance of Throughput-based User Association}

\begin{figure}
	\centering
	\includegraphics[width=0.9\linewidth]{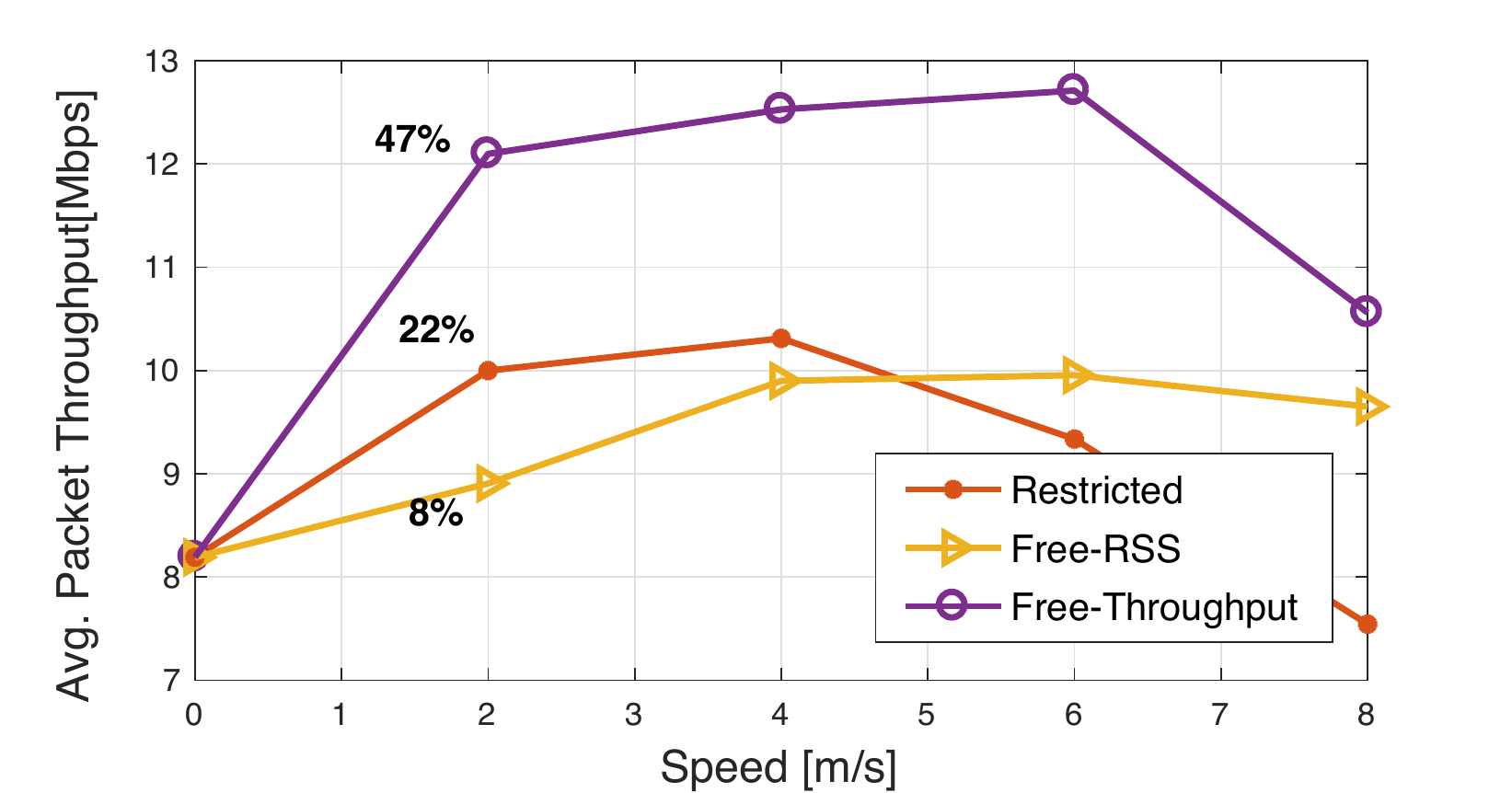}
	\caption{Average packet throughput of different movement models for acceleration of 4$m/s^2$}
	\label{fig:packetthall}
\end{figure}

\begin{figure}
	\centering
	\includegraphics[width=0.95\linewidth]{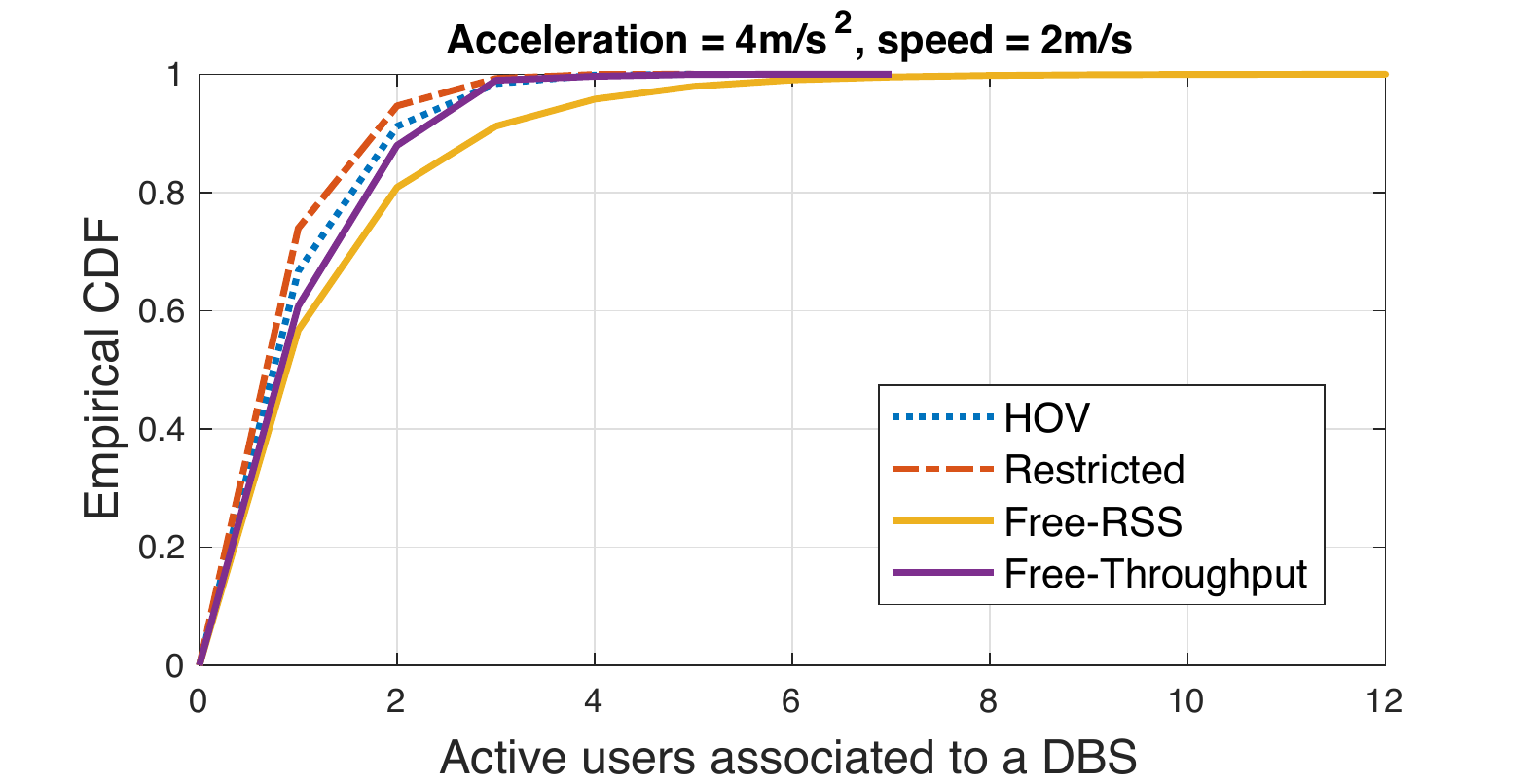}
	\caption{CDF of number of associated active users to DBSs }
	\label{fig:usersall}
\end{figure}

In this section, we study the performance of the more intelligent user association scheme (Throughput-based). Figure \ref{fig:packetthall} shows
that employing the Throughput-based user association scheme improves the packet throughput significantly. According to this figure, DBSs in the \textit{Free-Throughput} model with the acceleration value of 4$m/s^2$, and the speed of 2m/s achieves a remarkable packet throughout gain of 47\% compared to the \textit{HOV} scenario, while the achievable gain for \textit{Free-RSS} and \textit{Restricted} movement model are 8\% and 22\%, respectively.
Particularly, it shows that by having a smart user association scheme, freeing up the DBS in the network generates better results than limiting them to serve local users within a small cell boundaries. This huge improvement is the result of balanced load among DBSs, as shown in Figure \ref{fig:usersall}. This figure illustrates the distribution of number of active associated users to DBSs during the simulation time, when drones are moving with the speed of 2m/s. It shows that deploying Throughput-based scheme balances the loads as achieved by the \textit{Restricted} and \textit{HOV} movement model.

\subsection{The DBSs Collision Issue}

\begin{figure}
	\centering
	\includegraphics[width=0.9\linewidth]{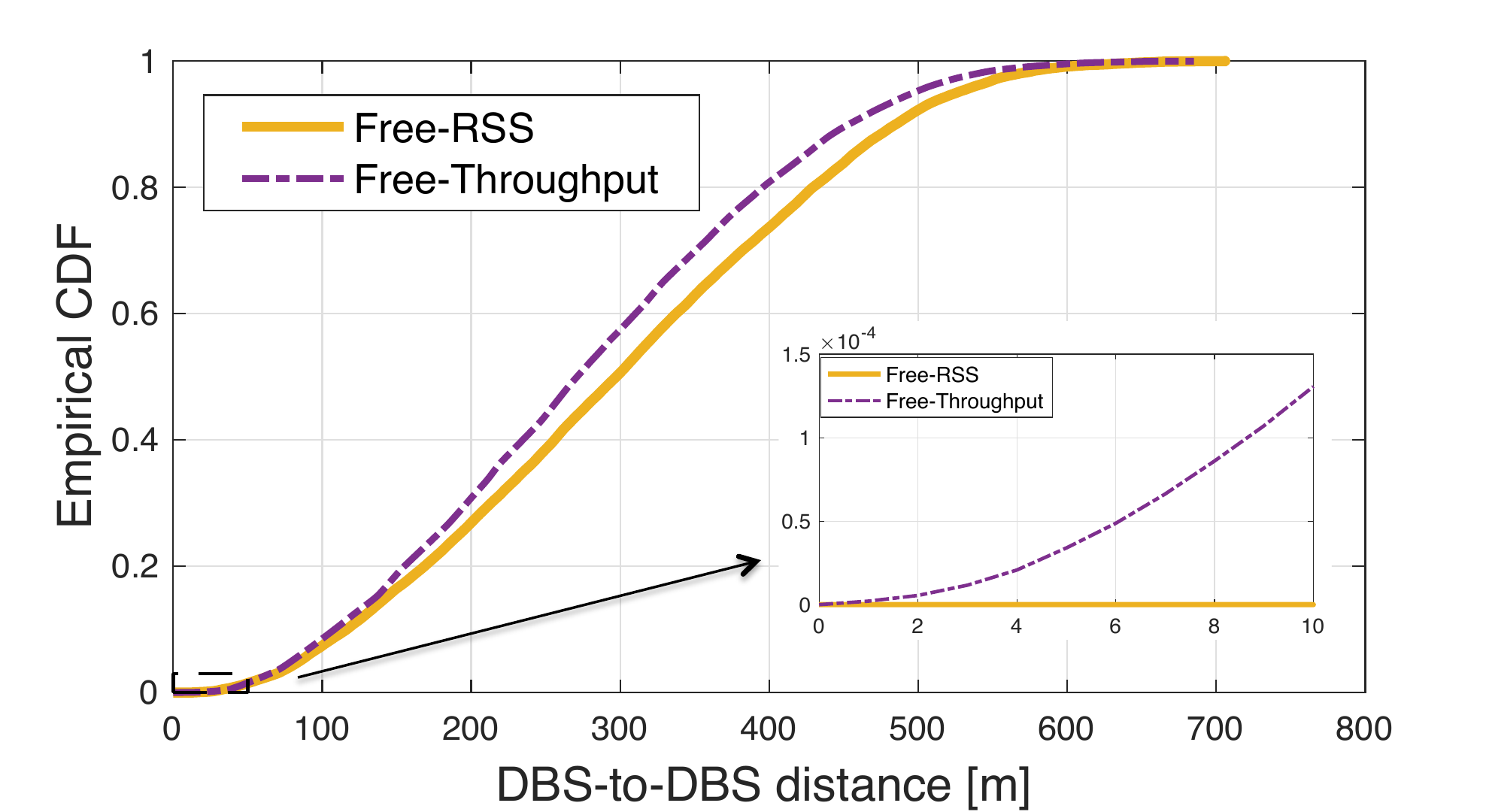}
	\caption{Empirical CDF of DBS-to-DBS distance}
	\label{fig:dbstodbsdistance}
\end{figure}

Note that when drones are moving freely at the same height, they may collide with each other. To study the probability of collision, we analyzed the DBS-to-DBS distance during the simulation time. Figure \ref{fig:dbstodbsdistance} illustrates the CDF of such DBS-to-DBS distance for the free movement models. As can be seen from this figure, the intelligent movement of drones maintains a comfortable distance among the DBSs. The intuition is that in the proposed DMA, each DBS tends to be closer to its serving users, and farther away from interfering DBSs. Therefore, the possibility of having two drones flying in close proximity is extremely low, which helps to prevent drones coming too close to each other. As shown in Figure \ref{fig:dbstodbsdistance} the probability that the DBS-to-DBS distance is less than 10m, is well below $0.00015$.

\section{Conclusion and Future Work} \label{sec:conclusion}
In this paper, we have shown that by freeing up the DBSs from the cells and letting them cruise in the network, a significantly larger system throughput can be achieved compared with the case that each DBS is restricted within each cell. However, this huge performance comes at the expense of an intelligent and complex user association scheme, which needs global network knowledge. Designing less complex user association scheme to gain benefits from the free movement model is left for future work. Moreover,
by allowing DBSs to move freely, the opportunity to deploy a less number of DBSs becomes promising. The performance of different number of DBSs in the network will be studied in future work.
\section*{Acknowledgment}

Azade's research is supported by Australian Government Research Training Program Scholarship and Data61$|$CSIRO PhD top-up scholarship.

\bibliographystyle{IEEEtran}
\bibliography{ws_gc_freemovement}

\end{document}